# Review of Machine-Learning Methods for RNA Secondary Structure Prediction

Qi Zhao[1#], Zheng Zhao[2#], Xiaoya Fan[3], Zhengwei Yuan[4], Qian Mao[5], Yudong Yao[6*]

1. College of Medicine and Biological Information Engineering, Northeastern University, Shenyang, Liaoning, 110169, China, zhaoqi@mail.neu.edu.cn

2. School of Information Science and Technology, Dalian Maritime University, Dalian, Liaoning, 116026, China, zhaozheng@dlmu.edu.cn

3. School of Software, Dalian University of Technology, Key Laboratory for Ubiquitous Network and Service Software of Liaoning Province, Dalian, Liaoning, 116620, China, xiaoyafan@dlut.edu.cn

4. Key Laboratory of Health Ministry for Congenital Malformation, Shengjing Hospital, China Medical University, Shenyang, Liaoning, 110004, China, yuanzw@hotmail.com

5. College of Light Industry, Liaoning University, Shenyang, Liaoning, 110036, China, qianmao72@163.com

6. Department of Electrical and Computer Engineering, Stevens Institute of Technology, Hoboken, NJ, 07030, United States, yyao@stevens.edu

* Corresponding Author. Email address: yyao@stevens.edu

# These authors contributed equally to this work.

**Abstract**

Secondary structure plays an important role in determining the function of non-coding RNAs. Hence, identifying RNA secondary structures is of great value to research. Computational prediction is a mainstream approach for predicting RNA secondary structure. Unfortunately, even though new methods have been proposed over the past 40 years, the performance of computational prediction methods has stagnated in the last decade. Recently, with the increasing availability of RNA structure data, new methods based on machine-learning technologies, especially deep learning, have alleviated the issue. In this review, we provide a comprehensive overview of RNA secondary structure prediction methods based on machine-learning technologies and a tabularized summary of the most important methods in this field. The current pending issues in the field of RNA secondary structure prediction and future trends are also discussed.

**Keywords** RNA secondary structure prediction; Machine learning; Deep learning

## 1. Introduction

Since its discovery, for a long time, RNA was regarded solely as a message carrier between DNA and protein. However, we are now beginning to understand its important roles, as increasing numbers of non-coding RNAs (ncRNA) are being discovered [1]. According to the latest report, less than 2% of the human genome comprises protein-coding regions [2]. Majority of the remaining genome portions encode ncRNAs [3], which are involved in translation, catalysis, RNA stability, RNA modification, gene expression regulation, protein synthesis and protein degradation [4-9]. The enormous importance of ncRNAs in various human diseases, such as cancer, diabetes, and atherosclerosis [6, 10], is also being recognized.

NcRNA molecules often fold into higher-order structures, and functionally-important ncRNA structures are typically conserved during evolution. Similar to protein, the ncRNA function is usually closely related to its structure. Currently, a wide variety of ncRNA sequences are publicly available and their numbers keep dramatically increasing [11]. By contrast, most of their structures remain unknown, which hinders the inference of their function. Hence, efficient determination of ncRNA structure is of great value to research.

Unlike the global folding of protein driven by hydrophobic forces, the RNA folding process is hierarchical (Fig. 1). Specifically, the RNA secondary structure, composed of base pairs, forms rapidly from linear RNA (primary structure) with a large energy lost, while the formation of a complex tertiary structure (or 3D structure) is usually much slower [12]. The RNA secondary structure is very stable and abundant in the cell, which is important for ncRNA function. Even without the knowledge of higher-order structure, RNA secondary structure is sufficient for a functional inference and other practical applications.

Computational predictions are mainstream approaches for identifying RNA secondary structure. A number of prediction methods have been developed since the 1970s. Most of these methods attempt to identify a structure with a minimum free energy (MFE), in agreement with the hypothesis that an RNA molecule is likely to exist in an MFE state, just like protein [13]. Many prominent software applications have been developed incorporating these methods [14-16]. However, in the last 10 years, the accuracy of prediction failed to significantly improve, neither did the calculating speed. An alternative approach, the machine-learning (ML)-based methodology, was proposed to improve the predictions of RNA secondary structure. However, such methods did not receive much attention because of the limited accuracy. That was mainly because a small size of the training datasets and the limitations of simple ML models. As a result of the recent explosion of RNA sequence data and the improvement of ML techniques, especially deep-learning (DL) techniques, the latest ML-based methods supersede the current mainstream methods in terms of accuracy and applicability. We believe that these ML-based methods will inspire the next generation of prediction tools in the near future.

In this paper, we provide a comprehensive overview of ML-based methods for RNA secondary structure prediction, with a thorough discussion of their advantages and disadvantages. We also provide a tabularized summary (Table 1) of the most important models in the field, and a perspective on the future promising directions, with a special emphasis on DL models. Although several review papers have been published on the topic of RNA secondary structure prediction [17-19], reviews with an emphasis on ML techniques are lacking. We believe that this review will enable researchers to understand the key issues that remain to be solved and facilitate further advances in predicting the RNA secondary structures based on ML.

**2. RNA secondary structure: the basics**

The RNA molecule is an ordered sequence of nucleotides that contain one of the four bases: adenine (A), cytosine (C), guanine (G), and uracil (U), arranged in the 5' to 3' direction. Pairing (via hydrogen bonds) of these four bases within an RNA molecule gives rise to the secondary structure. Typically, each base pairs with at most one other base. The canonical base pairs include the Watson-Crick base pairs (A-U and G-C) and the wobble base pair (G-U). These base pairs often result in the formation of a nested structure, in which multiple stacked base pairs form a helix, and one or multiple unpaired base pairs form a loop.

In addition, a small number of special base pairs [20] exist in the native RNA secondary structure, including non-canonical base pairs (not A-U, G-C, and G-U pairs) and base triplets (with base-pair formation between three bases at once). Another type of a special base pair is a pseudoknot, which occurs when bases in different loops pair with each other. Pseudoknots form non-nested structures between two bases that are located apart from each other, and often play an important role in RNA function [21]. Overall, the special base pairs account for 10% of all base pairs in the native molecule.

Typically, the secondary structure of an RNA molecule with a length $n$ can be regarded as:

1) A set of base pairs $\{(i,j), 1 \leq i < j \leq n\}$, where $(i,j)$ indicates a base pair formed between the $i$-th and $j$-th nucleotide in the RNA sequence; or a set of compatible helixes [22].

2) A contact table, *i.e.*, a square matrix, with elements in the $i$-th row and $j$-th column representing the interaction between the $i$-th and $j$-th nucleotides in the RNA sequence.

3) A graph, where nodes represent nucleotides and edges represent base pairs.

4) A labeled sequence with the length $n$, *e.g.*, "dot-parenthesis" notation, with matching parentheses for paired bases and dots for unpaired bases.

5) A parse tree derived from context-free grammars, of which the leaf nodes comprise the RNA sequence [23].

The above definitions are the bases of both traditional and ML-based RNA secondary structure prediction methods.

### 3. Traditional methods of detecting or predicting RNA secondary structure

RNA structure determination is a fast-evolving topic. Many different methods have emerged in the last 20 years. They can be divided into two categories, *i.e.*, wet-lab experimental approaches and computational predicting approaches.

### 3.1 Wet-lab experiments

X-ray crystallography [24] and nuclear magnetic resonance (NMR) [25] are the most accurate approaches for determining RNA structure, both of which can offer structural information at a single base-pair resolution. However, both methods are characterized by high experimental cost and low throughput, limiting their wide usage. In addition, RNA molecules are highly unstable and difficult to crystallize. Although many methods have been developed to infer the state of nucleotides (paired or unpaired) in an RNA molecule using enzymatic [26, 27] or chemical probes [28, 29] coupled with next-generation sequencing [30, 31], most of them can only be used to capture the RNA secondary structure *in vitro*. The obtained structure may differ markedly from the *in vivo* conformation. In fact, to date, the structure of only a very small percentage (< 0.001%) of known ncRNAs has been determined experimentally [32]. Hence, predicting the RNA secondary structure using computational methods is an important alternative to wet-lab–based approaches.

**3.2 Traditional computational methods**

Comparative sequence analysis [33, 34] is the most accurate computational method for determining the RNA secondary structure. This method is based on the principle that the RNA secondary structure is evolutionarily conserved to a greater extent than its corresponding sequence. However, while this method requires a large set of homologous sequences aligned manually by experts, only thousands of RNA families are currently known [35]. This makes it impossible to obtain homologous sequences for any specific RNA. Therefore, most methods for RNA secondary structure prediction focus on a score-based method, which requires a single RNA sequence as input.

Score-based methods are the most widely used methods and have dominated the field of RNA secondary structure prediction in the last four decades. These methods assume that the native RNA structure is a structure with a minimum/maximum total score, depending on the hypothesis of RNA folding mechanism or its simplification. Hence, the problem of RNA secondary structure prediction is transformed into an optimization problem. Since the RNA secondary structure can be recursively broken down into smaller elements with independent score contributions, dynamic programming (DP) algorithm is often employed to identify the optimal structure. Evaluation of the score for structure elements requires a score scheme of many parameters. Nussinov and Jacobson [36] proposed the first, and also the simplest, DP algorithm for finding the maximum-matching structure. The authors proposed to assign 1 point to each matched base pair and assumed that the native structure is the structure with the maximum score among all the possible conformations. Zuker *et al*. [37] proposed a more realistic scoring scheme based on free energy, the nearest-neighbor model (NN model) [38-41]. It is based on Tinoco's hypothesis [42] (See Section 4.1). The NN model can be used for the calculation of energy changes of any structure of a given RNA molecule, and DP algorithm was also employed to efficiently find the MFE structure. A number of slightly different variations of this method were also proposed [43-46].

However, the folding mechanism hypotheses of score-based methods do not always hold, *e.g.*, the RNA molecule often folds into locally stable structural domains. Furthermore, almost all score-based methods use virtually the same DP algorithm to find the best-scoring structures. However, the running time of the DP algorithm is usually $O(n^3)$ (where $n$ is the RNA sequence length), neglecting the special base pairs and weak interactions. Hence, the computational cost is not acceptable, especially when analyzing an RNA molecule that is more than 1000 nucleotides long.

In fact, it is extremely difficult to fully understand the RNA folding mechanism. ML methods, in contrast, are data-driven, and requiring no knowledge of such mechanism. These methods can learn the underlying patterns in a dataset. In the last few decades, ML methods have been used for many aspects of RNA secondary structure prediction methods to improve the prediction performance (See Section 4). However, they did not replace the mainstream score-based methods with respect to accuracy and generalization. This situation has been changing in the last 2 years because of the development of ML techniques, especially DL.

**4. ML-based methods**

The ML-based methods for RNA secondary structure prediction can generally be divided into three categories (Fig. 2), according to the sub-process that ML participates in: score schemes based on the ML model; preprocessing and postprocessing based on the ML model; and a prediction process based on the ML model.

**4.1 Score schemes based on the ML model**

According to the meaning of the score, ML-based score schemes (Fig. 3) can be further divided into three categories, *i.e.*, the free-energy parameter refining approach, weighted approach, and probabilistic approach. Although ML-based methods are used for parameter estimation in the score schemes to improve the prediction accuracy, the structure prediction is still formulated as an optimization problem, where the estimated parameters are used for the evaluation of scores of possible conformations.

**1) Free-energy parameter refining based on ML**

Considering the score schemes, the free-energy-focused approach is the most popular approach. Ever since Tinoco *et al*. [42] put forward their hypothesis for free-energy calculation (that the free energy of a secondary structure is the sum of the free energy values of its elements), many researches have been devoting their efforts to the problem of assigning free-energy values to the elements of RNA molecules. Turner's NN model [41] is the most popular approach, and provides a considerably accurate approximation of the RNA free energy. However, the multiple thermodynamic parameters of the NN model have to be based on a large number of optimal melting experiments. These experiments are time- and labor-consuming [14, 16], however, and not all free-energy changes in structural elements can be measured because of the associated technical difficulties.

Some ML techniques were adopted to refine the parameters in the energy model. These techniques can employ subtle models to estimate the scores for a more rich and accurate feature representation using known thermodynamic data/RNA secondary structure data. Xia *et al*. [40] first trained a linear regression model using known thermodynamic data to infer some of the thermodynamic parameters, and expanded the NN model into a more accurate model, *i.e.*, the INN-HB model. This model provides an improved fit for the known experimental data. A disadvantage of this approach, however, is that the parameters for some structural elements are fixed before other parameters are calculated, which limits the range of possibilities considered for the overall parameter set. To overcome this problem, Andronescu *et al*. [47] proposed a constraint-generation approach to estimate free-energy parameters. This method uses different types of constraints to ensure that the energies of reference structures are low relative to the alternatives for the same sequence. Trained on large sets of structural and thermodynamic data, this method achieves 7% higher F-measure than the standard Turner parameters. Subsequently, the authors further modified the method, and proposed a loss-augmented max-margin constraint-generation model and Boltzmann-likelihood model using a larger dataset [48]. The constraints imposed on parameters ensure that the more inaccurate the structure, the greater the margin between its free energy and that of the reference structure in the training set.

Of note, the parameters determined by the above free-energy parameter-refining approaches are thermodynamic, and can be used directly in the algorithms embedded by the same energy model, such as miRNA target prediction [49] and RNA folding kinetics simulation [50].

## 2) Weighted approaches based on ML

While ML-based free-energy parameter approaches successfully improved the accuracy of the RNA secondary structure prediction, the energy model is still far from ideal. Actually, the above methods for the estimation of ML-based parameters can only substitute for some wet-lab experiments geared toward obtaining the energy parameters. However, it is entirely possible to obtain an improved score scheme independent of free energy, based on ML techniques. Several weighted approaches were proposed that consider the parameters of RNA structure elements as weights instead of free-energy changes. Getting rid of the thermodynamic meaning, the weighted approach can utilize ML models to determine thousands of weights for more comprehensive RNA structure elements instead of obtaining few energy parameters from a large number of wet-lab experiments.

By combining a discriminative structured-prediction learning framework with an online learning algorithm, Zakov *et al*. [51] greatly increased the number of weights to approximately 70,000 by examining more types of structural elements with more numerous sequential contexts, using thousands of training datasets. Based on these weights, ContextFold tool was proposed, marking a significant improvement in the prediction accuracy [51]. However, as reported by Rivas *et al*. [52], ContextFold may suffer from over-fitting. Akiyama *et al.* [53] integrated the thermodynamic approach with a structured support vector machine (SSVM) to obtain a large number of weights for detailed structure elements, of which $l1$ regularization was used to relieve over-fitting. Then, mx-fold was built by combining ML-based weights with experimentally-determined thermodynamic parameters, achieving better performance than a model based on thermodynamic parameters or ML-based weights alone. This suggests that ML-based weights can complement the gaps in the thermodynamic parameter approach.

An advantage of the weighted approach is that it decouples structure prediction from energy estimation, which is potentially beneficial for both tasks. However, learned weights are not explainable, partly because of the black-box attribute of ML algorithms. Hence, the obtained scores cannot be used to compute the partition function, base-pair binding probabilities, or centroid structures, *etc*.

## 3) Probabilistic approaches based on ML

Statistic context-free grammars (SCFGs) are an important alternative for predicting RNA structure [23, 54-58]. SCFGs specify formal grammar rules and induce a joint probability distribution over possible RNA structures for a given sequence. In particular, an SCFG model specifies a probability parameter for each production rule in the grammar, and thus assigns a probability to each sequence it derives. The probability parameters are learned from datasets of RNA sequences annotated using known secondary structures, without the need for external laboratory experiments [57].

Sakakibara *et al*. [23] first applied SCFGs to tRNA secondary structure prediction. The probability parameters in their SCFG model were learned using an expectation-maximization (EM) method. Knudsen and Hein [56] improved the SCFG model by combining the evolutionary information and subsequently, the robust and practical tool Pfold [55] was developed. Sato *et al*. [59] proposed a non-parametric Bayesian extension of SCFGs with the hierarchical Dirichlet process to find an optimal RNA grammar from the training dataset. Using another ML model, the conditional log-linear model (CLLM), Do *et al*. [60] identified probability parameters that are most likely to discriminate correct structures from incorrect ones. CLLM is a flexible probabilistic ML model that generalizes upon SCFGs; the parameters are easily estimated and arbitrary features can be incorporated in the model. CONTRAfold has achieved the highest single-sequence prediction accuracy to date, compared with the currently available probabilistic models. However, CLLM is very slow, which prevents its application to large training sets, and the estimated parameters have no intrinsic biological meaning. Finally, to take full advantage of the substantial numbers of RNA sequences with unknown structures, Yonemoto *et al.* [61] proposed a semi-supervised learning algorithm to obtain probability parameters in a probabilistic model that combines SCFG and a conditional random field.

However, the probabilistic approach cannot replace MFE methods for secondary structure prediction, as the accuracy of the currently best SCFG has yet to match those of the best free-energy-based models. In addition, SCFG cannot describe all RNA structures, *e.g.*, a structure containing special base pairs.

### 4.2 Preprocessing and postprocessing based on ML model

ML can be also used in pretreatment, for selecting an appropriate prediction method or a group of appropriate parameters (Fig. 4). A tool based on a support vector machine was proposed by Hor *et al*. [62] for selecting the prediction method, based on the notion that different RNA sequences have different features and different prediction methods work best with specific RNA species. In another study, Zhu *et al*. [63] assumed that different RNA sequences follow different folding rules. The authors consequently proposed an SCFG model to identify the most probable folding rules before RNA secondary structure prediction.

Since different prediction methods return several different structures, the ML model can provide a means of determining the most likely structures among the outcomes (Fig. 4). Combined with the graph theory, Haynes *et al*. [64] used trees to represent RNA graphical structures (edges as helices, and verticals as loops or bulges). They then trained a multi-layer perceptron (MLP) model to distinguish whether a structure is RNA-like or not, using graphical invariants as input features. Assuming that a larger secondary structure is formed upon bonding of two smaller secondary RNA structures, Koessler *et al*. [65] also used an MLP model to predict the RNA-like probability of a structure using a special feature-vector extracting from the merged trees [65].

### 4.3 Predicting processes based on ML model

ML techniques are also directly used to predict RNA secondary structure in an end-to-end fashion, or combined with other algorithms as constrains, base-state detector, or structure selector. The general framework is shown in Fig. 5.

**1) End-to-end approach**

To the best of our knowledge, the ML technique was first introduced into the RNA secondary structure predicting process by Takefuji *et al.* [66]. The authors built on the Nussinov and Jacobson's hypothesis (See Section 3.2) [36], and attempted to obtain a near-maximum independent set (MIS) from an adjacent graph (where the vertices represent base pairs, and the edges connect the incompatible vertices) using a system composed of $m$ interactional neurons ($m$ is the number of edges). Liu *et al*. [67] enhanced Takefuji's work by considering the energy contribution of possible base pairs, and a Hopfield neural network (HNN) was employed to obtain MIS. However, HNN is easily trapped in local minima, limiting the accuracy of this method. To avoid this problem, Steeg [68] made use of the Mean Field Theory (MFT) networks to identify the optimal structure, which was coupled with a sophisticated objective function with additional biological constrains. The inputs into the MFT networks are the four types of bases in an RNA sequence encoded in a one-hot fashion, and the output is in a format similar to contact-table. Subsequently, Apolloni *et al*. [69] further developed the Steeg's method, especially with respect to the computation speed, so that it could be applied to slightly longer RNA sequences. In addition, this model uses mean-field approximation to update the node in both, the learning phase and the instant-resolution phase. In another study, Qasim *et al*. [70] modified Takefuji's work by building a novel MLP model to obtain MIS. This model contains $h$ neurons in the hidden layer, whose activate function is based on the Kolgomorov's theorem ($h$ is the number of possible base pairs in an RNA sequence).

However, because of the relatively poor performance of the above ML models and a small amount of the available data, ML-based RNA secondary structure prediction models can only process tRNAs, with relatively low accuracy. Currently, the use of DL techniques is rising rapidly, and they are dramatically changing such circumstance. Singh *et al*. [71] proposed the first end-to-end DL model, SPOT-RNA, to predict RNA secondary structure. SPOT-RNA treats the RNA secondary structure as a contact table and employs an ensemble of ultra-deep hybrid networks of ResNets and 2D-BLSTMs for the prediction. Of these, the former captures the contextual information from the whole sequence, and the latter is effective for the propagation of long-range sequence dependencies in RNA structure. Transfer learning is used to train SPOT-RNA to effectively utilize limited sample numbers. E2Efold is another successful DL model for RNA secondary structure prediction, proposed by Chen *et al*. [72]. It integrates two coupled parts: a transformer-based deep model that represents sequence information, and a multi-layer network based on an unrolled algorithm that gradually enforces the constraints and restricts the output space. Both SPOT-RNA and E2Efold showed superior performance with several RNA benchmark datasets, greatly outperforming the best score-based methods and SCFG-based methods.

In addition to the encoded RNA sequences being used as the input, other information can also be incorporated into the DL model. Calonaci *et al*. [73] trained an ensemble model based on a combination of SHAPE data, co-evolutionary data (DCA), and RNA sequence. Their model consists of a CNN sub-network and an MLP sub-network to predict penalty based on SHAPE and DCA data, respectively, with an RNAfold [14] module to generate structure using RNA sequence and penalties.

**2) Hybrid approach**

Alternatively, ML can be combined with other methods for a hybrid approach for RNA secondary structure prediction. Consequently, the ML model is usually considered as a scoring machine, mapping a score to each (pair of) base(s) in an RNA sequence, whose output is then passed to an independent filter to identify a reasonable structure.

Bindewald *et al.* [74] combined an ML model and a filter to predict the consensus structure for a group of aligned RNAs. The authors chose a hierarchical network of k-nearest neighbor model to predict the possibility score for each pair of alignment columns, and defined the filter by a set of rules derived from native RNA structures. Considering structure prediction as a sequence-labeling question, Lu *et al.* [75] and Wu *et al.* [76] employed a more powerful DL model, Bi-LSTM, to predict the state of each base in an RNA sequence, using a similar rule-based filter to deal with conflicting pairing. Differently from the above studies, Bi-LSTM was used as a structure filter in DpacoRNA [77], and a parallel ant-colony optimization method was used to predict the most probable structures. Another type of an ML-based hybrid approach combines ML models and optimization methods. Liu's group [78] used a CNN model to predict the status distribution of each base in an RNA sequence, and a DP algorithm was employed to find the maximum-probability structure. The same group [79] also used the Bi-LSTM model instead and another optimization algorithm, similar to that used in [22]. Instead of developing a new optimizer, Willmott *et al.* [80] utilized an existing SHAPE-directed method (SDM) [81] as the optimizer, which can predict optimal structure from SHAPE data, and trained a Bi-LSTM model to generate SHAPE-like data (*i.e.*, determine the state of each nucleotide) of an RNA sequence as the inputs of SDM.

Compared with the end-to-end approach, the performance of the hybrid approach is relatively poor, perhaps because of a bias between the training objective of the ML part and the overall system objective. Most methods in the hybrid approach are trained and tested using small-scale datasets. Hence, generalization of their abilities requires further verification.

## 5. Perspectives

It is well known that transcript abundance helps to identify transcripts of interest under different conditions, while the RNA structure helps to explain how these transcripts function. An excellent RNA structure prediction method is not only important for inferring RNA function, but also relates to many downstream studie*s*, including ncRNA detection [82-84], folding dynamics simulations [85], hybridization stability assessment [86], and oligonucleotide [87, 88] or drug design [89-93]. It is worth noting that RNA secondary structure prediction is also a useful tool for studying viruses, such as the SARS-CoV-2 virus responsible for the current pandemic [94, 95].

### 5.1 The advantages of ML-based methods

Compared with comparative sequence analysis and traditional score-based methods, ML-based methods have some advantages. First, ML-based methods do not necessarily rely on the biological mechanism, which is usually difficult to thoroughly understand. Instead, they can utilize the information contained in various types of data and, therefore, performance limitation caused by the mechanism hypothesis can be circumvented. ML-based methods can also be easily coupled with known biological mechanisms. Further, in terms of prediction performance, where a large amount of data is available, models with the no/less knowledge of biological mechanisms usually perform better than mechanism-depended ones. This also suggests that the assumed mechanism of RNA folding may be incomplete or not accurate. Second, ML-based methods can be considerably flexible, which gives a possibility for achieving better results. The inputs of ML-based models can be either one-dimensional or multi-dimensional, extracted features or encoded bases, and homogeneous data or heterogeneous data; and the outputs can be contact tables, labeled sequences, nucleotide states, or free energy values. In addition, the construction of the ML models is diverse, from simple Hopfield networks to complex ensemble deep-neural networks. Third, once the model training is completed, the ML-based end-to-end prediction methods run very fast. Unlike DP algorithm, the time complexity of ML models is independent of the input scale, which is advantageous when dealing with long RNAs.

### 5.2 Current pending issues

Enormous progress has been made toward predicting RNA secondary structure by using ML-based methods. These methods are state-of-the-art when considering most indices of prediction performance. However, some issues still require resolving.

First, the accuracy of prediction should be further improved, especially when some special base pairs exist in the native RNA structure. In fact, many traditional methods neglect special base pairs to avoid a large number of false-positives or limiting computational complexity [96, 97]. While some methods can predict RNA secondary structures containing pseudoknots [98] or non-canonical base pairs [99], none of them can predict both. Although the recently proposed ML-based methods can predict all kinds of special base pairs, the prediction accuracy is still limited [71].

The RNA sequence length limitation is another intractable issue, which becomes quite problematic with the recently discovered long (1000–10,000 nt) ncRNA [100]. Although ML-based methods do not suffer from high-time complexity as most other score-based methods, they are unable to effectively capture such long-range interactions within an RNA sequence. On the other hand, training an ML model with such a large number of inputs consumes a huge amount of computational resources, and is often unrealistic.

Generally, the enhancement of predictive ability is associated with the relatively large scale of the ML model, which requires large amounts of data for parameter training. Although tens of thousands of RNA structure data in various formats are available in databases, such as RNAstrand [101], bpRNA [102], RAG [103], and Rfam [35], these are insufficient, especially with respect to high-accuracy data availability, in terms of training large-scale DL models. Hence, questions on how to effectively utilize the limited data and cope with over-fitting of a large DL model are also important issues that remain to be resolved.

**5.3 Future trends of development**

Currently, RNA secondary structure prediction is successfully shifting toward ML-based approaches, away from traditional score-based methods, and DL will surely continue to improve the prediction performance. The subtle structure of the DL model is a prerequisite to this end. Since the DL model is rapidly developed in natural language processing and image processing field, using mature models or combining them in such fields constitutes a feasible way for generating an excellent DL model for RNA secondary structure prediction.

Further, using a DL model to predict the free-energy parameter is an inevitable trend for more accurate energy estimations, when additional wet-lab experimental data become available. However, these parameters may not improve RNA secondary prediction accuracy because they have to be combined with traditional MFE-based methods. On the other hand, combing an ML-based method and an optimization method is a promising approach for improving prediction performance.

**6. Conclusion**

RNA structure is one of the central pieces of information for understanding biological processes, and determining RNA secondary structure will continue to be a hot topic in the computation and biology fields. In this review, we focused on ML-based methods, which involve many aspects of RNA secondary structure prediction. ML techniques have greatly improved the performance of prediction methods, including accuracy, applicability, and running speed. However, to thoroughly resolve the RNA secondary structure prediction problem, a more subtle ML model is still needed. At the moment, ML based-methods cannot be used as substitutes for wet-lab experiments for obtaining high-resolution structures. Nonetheless, the advent of DL technologies and high-performance hardware will foster a new generation of RNA secondary prediction tools with an improved accuracy and running speed.

**Funding**

This work was supported in part by the Fundamental Research Funds of Northeastern University (N181903008); the Research Start-up Fund for Talent of Dalian Maritime University (02500348); National Natural Science Foundation of China (31801623); the Fundamental Research Funds for the Central Universities (82232019); the National Natural Science Foundation of China (81871219).

**Competing interests**

The authors declare that they have no competing interests.

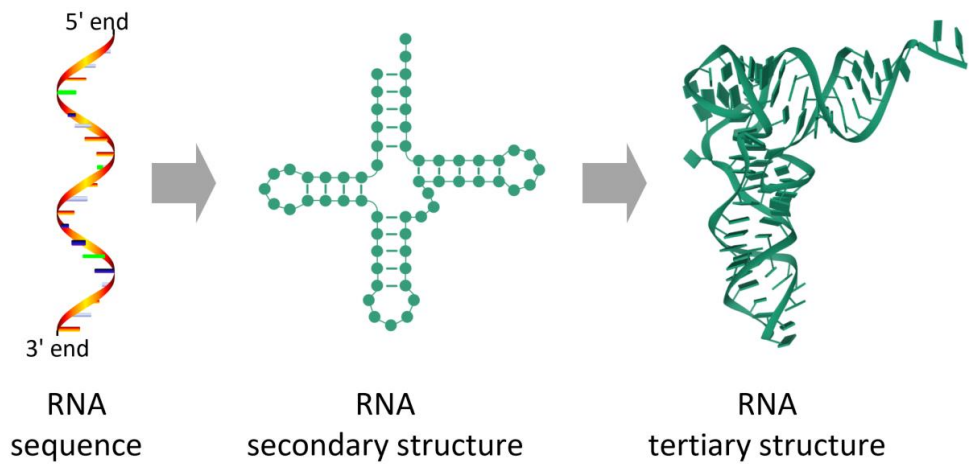

**Fig. 1. RNA primary, secondary, and tertiary structures**

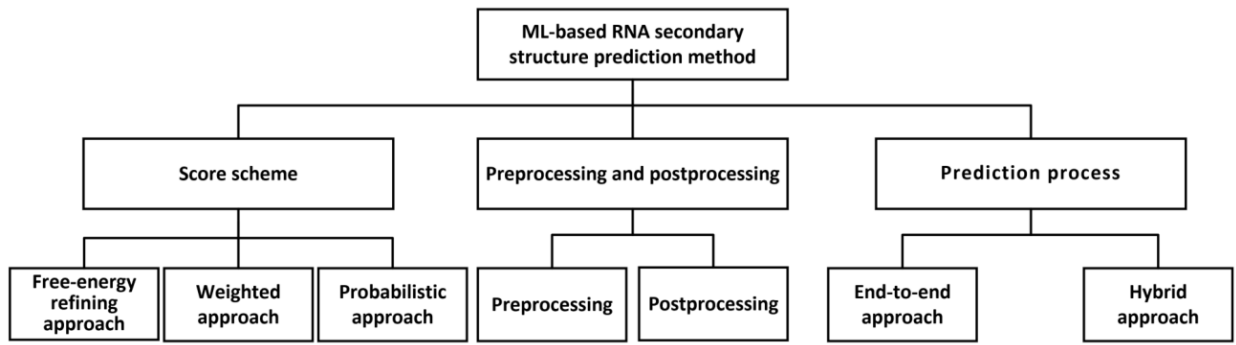

Fig. 2. Classification of ML-based RNA secondary structure prediction methods

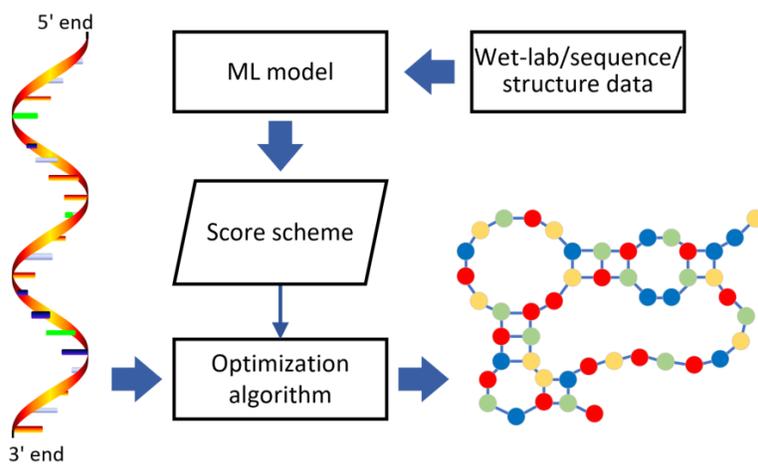

**Fig. 3. Framework for RNA secondary structure prediction methods with an ML-based score scheme**

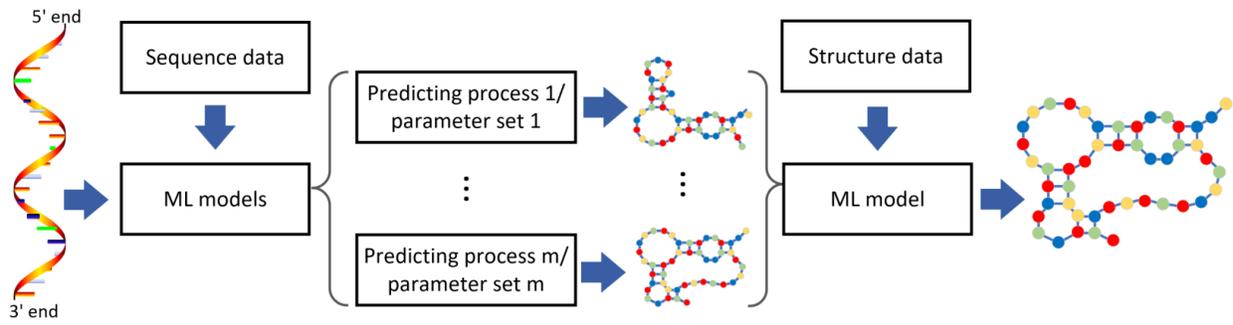

**Fig. 4.** Framework for RNA secondary structure prediction methods with ML-based preprocessing or postprocessing

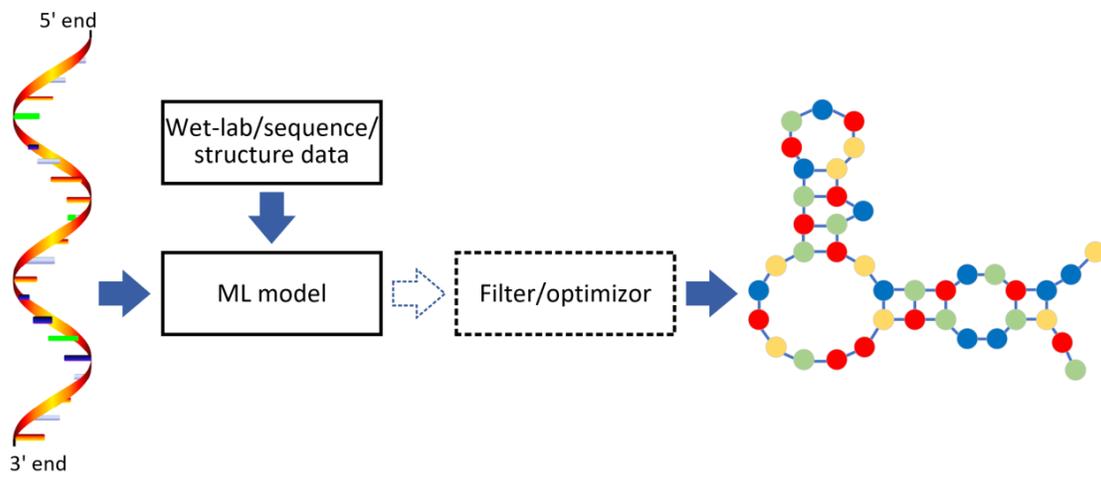

**Fig. 5. Framework for the RNA secondary structure prediction methods with ML-based predicting process**

**Table 1. Summary of the ML-based RNA secondary structure prediction methods**

| Category | | Title | Date | Author | ML Technique | Reference |
|---|---|---|---|---|---|---|
| Score schemes based on ML model | Free-energy parameter-refining approach based on ML | Thermodynamic Parameters for an Expanded Nearest-Neighbor Model for Formation of RNA Duplexes with Watson-Crick Base Pairs | 1998 | Xia *et al.* | Linear regression | [40] |
| | | Efficient parameter estimation for RNA secondary structure prediction | 2007 | Andronescu *et al.* | Constraint generation | [47] |
| | | Computational approaches for RNA energy parameter estimation | 2010 | Andronescu *et al.* | Loss-augmented Max-margin Constraint Generation model, Boltzmann-likelihood model | [48] |
| | Weighted approach based on ML | Rich Parameterization Improves RNA Structure Prediction | 2011 | Zakov *et al.* | discriminative structured-prediction learning framework combined, online learning algorithm | [51] |
| | | A Max-Margin Training of RNA Secondary Structure Prediction Integrated with the Thermodynamic Model | 2018 | Akiyama *et al.* | Structured support vector machine | [53] |
| | Probabilistic approach based on ML | Stochastic context-free grammars for tRNA modeling | 1994 | Sakakibara *et al.* | Expectation maximization method | [23] |
| | | RNA secondary structure prediction using stochastic context-free grammars and evolutionary history | 1999 | Knudsen and Hein | Expectation maximization method | [56] |
| | | Pfold: RNA secondary structure prediction using stochastic context-free grammars | 2003 | Knudsen and Hein | Expectation maximization method | [55] |
| | | CONTRAfold: RNA secondary structure prediction without physics-based models | 2006 | Do *et al.* | Conditional log-linear model | [60] |
| | | A semi-supervised learning approach for RNA secondary structure prediction | 2015 | Yonemoto *et al.* | Semi-supervised learning algorithm | [61] |
| Preprocessing and postprocessing based on ML | Preprocessing based on ML model | A tool preference choice method for RNA secondary structure prediction by SVM with statistical tests | 2013 | Hor *et al.* | Structured support vector machine | [62] |

| | | Research on folding diversity in statistical learning methods for RNA secondary structure prediction | 2018 | Zhu et al. | Statistical Context-free Grammar model | [63] |
|---|---|---|---|---|---|---|
| model | Postprocessing based on ML model | Using a neural network to identify secondary RNA structures quantified by graphical invariants | 2008 | Haynes et al. | Multi-layer perceptron | [64] |
| | | A predictive model for secondary RNA structure using graph theory and a neural network | 2010 | Koessler et al. | Multi-layer perceptron | [65] |
| Predicting process based on ML model | End-to-end approach | Parallel algorithms for finding a near-maximum independent set of a circle graph | 1990 | Takefuji et al. | System composed of several interactional neurons | [66] |
| | | A Hopfield Neural Network Based Algorithm for RNA Secondary Structure Prediction | 2006 | Liu et al. | Hopfield networks | [67] |
| | | Secondary Structure Prediction of RNA using Machine Learning Method | 2011 | Qasim et al. | Multi-layer perceptron | [70] |
| | | Neural Networks, Adaptive Optimization, and RNA Secondary Structure Prediction | 1993 | Steeg | Mean Field Theory network | [68] |
| | | RNA secondary structure prediction by MFT neural networks | 2003 | Apolloni et al. | Mean Field Theory network with mean field approximation to update network's nodes | [69] |
| | | RNA secondary structure prediction using an ensemble of two-dimensional deep neural networks and transfer learning | 2019 | Singh et al. | Compound deep neural networks, transfer learning | [71] |
| | | RNA secondary structure prediction by learning unrolled algorithms | 2020 | Chen et al. | Compound deep neural networks | [72] |
| | | Machine learning a model for RNA structure prediction | 2020 | Calonaci et al. | Convolutional neural network, multi-layer perceptron | [73] |
| | Hybrid approach | RNA secondary structure prediction from sequence alignments using a network of k-nearest neighbor classifiers | 2006 | Bindewald et al. | Hierarchical network of k-nearest neighbor model | [74] |
| | | Developing parallel ant colonies filtered by deep learned constrains for | 2020 | Quan et al. | Bi-LSTM | [77] |

| | | predicting RNA secondary structure with pseudo-knots | | | | |
|---|---|---|---|---|---|---|
| | | RNA Secondary Structure Prediction Based on Long Short-Term Memory Model | 2018 | Yu *et al.* | Bi-LSTM | [76] |
| | | Predicting RNA secondary structure via adaptive deep recurrent neural networks with energy-based filter | 2019 | Lu *et al.* | Bi-LSTM | [75] |
| | | A New Method of RNA Secondary Structure Prediction Based on Convolutional Neural Network and Dynamic Programming | 2019 | Zhang *et al.* | Convolutional neural network | [78] |
| | | DMfold: A Novel Method to Predict RNA Secondary Structure with Pseudoknots Based on Deep Learning and Improved Base Pair Maximization Principle | 2019 | Wang *et al.* | Bi-LSTM | [79] |
| | | Improving RNA secondary structure prediction via state inference with deep recurrent neural networks | 2020 | Willmott *et al.* | Bi-LSTM | [80] |